# Deformation mechanisms and compressive response of NbTaTiZr alloy via machine learning potentials[*]

LIU Hongyang[1,2,3],　　CHEN Bo[1,2,3] ,　　CHEN Rong[1],　　KANG Dongdong[1,2,3],　　DAI Jiayu[1,2,3]

1.College of Science, National University of Defense Technology, Changsha 410073, China

2.Hunan Key Laboratory of Extreme Matter and Applications, National University of Defense Technology, Changsha 410073, China

3.Hunan Research Center of the Basic Discipline for Physical States, National University of Defense Technology, Changsha 410073, China

**Abstract**

Refractory multi-principal element alloys (RMPEAs) have become a hotspot in materials science research in recent years due to their excellent high-temperature mechanical properties and broad application prospects. However, the unique deformation mechanisms and mechanical behaviors of the NbTaTiZr quaternary RMPEA under extreme conditions such as high temperature and high strain rate are still unclear, limiting its further design and engineering applications. In order to reveal in depth the dynamic response of this alloy on an atomic scale, this study develops a high-accuracy machine learning potential (MLP) for the NbTaTiZr quaternary alloy and combines it with large-scale molecular dynamics (MD) simulations to systematically investigate the effects of crystallographic orientation, strain rate, temperature, and chemical composition on the mechanical properties and microstructural evolution mechanisms of the alloy under compressive loading. The results show that the NbTaTiZr alloy exhibits significant mechanical and structural anisotropy during uniaxial compression. The alloy exhibits the highest yield strength when loaded along the [111] crystallographic direction, while it shows the lowest yield strength when compressed along the [110] direction, where twinning is more likely to occur. Under compression along the [100] direction, the primary deformation mechanisms include local disordering transitions and dislocation slip, with $1/2\langle 111\rangle$ dislocations being the dominant type. When the strain rate increases to $10^{10}$ $s^{-1}$, the yield strength of the alloy is significantly enhanced, accompanied by a notable increase in the proportion of amorphous or disordered structures, indicating that high strain rate loading suppresses dislocation nucleation and motion while promoting disordering transitions. Simulations at varying temperatures indicate that the alloy maintains a high strength level even at temperatures as high as 2100 K. Compositional analysis further indicates that increasing the atomic percentage of Nb or Ta effectively enhances the yield strength of the alloy, whereas an increase in Ti or Zr content adversely affects the strength. By combining MLP with MD methods, this study elucidates the anisotropic characteristics of the mechanical behavior and the strain rate dependence of disordering transitions in the NbTaTiZr RMPEA under combination of high strain rate and high temperature, providing an important theoretical basis and simulation foundation for optimizing and designing novel material under extreme environments.

Refractory multi-principal element alloys (RMPEAs) have become a hotspot in materials science research in recent years due to their excellent high-temperature mechanical properties and broad application prospects. However, the unique deformation mechanisms and mechanical behaviors of the NbTaTiZr quaternary RMPEA under extreme conditions such as high temperature and high strain rate are still unclear, limiting its further design and engineering applications. In order to reveal in depth the dynamic response of this alloy on an atomic scale, this study develops a high-accuracy machine learning potential (MLP) for the NbTaTiZr quaternary alloy and combines it with large-scale molecular dynamics (MD) simulations to systematically investigate the effects of crystallographic orientation, strain rate, temperature, and chemical composition on the mechanical properties and microstructural evolution mechanisms of the alloy under compressive loading. The results show that the NbTaTiZr alloy exhibits significant mechanical and structural anisotropy during uniaxial compression. The alloy exhibits the highest yield strength when loaded along the [111] crystallographic direction, while it shows the lowest yield strength when compressed along the [110] direction, where twinning is more likely to occur. Under compression along the [100] direction, the primary deformation mechanisms include local disordering transitions and dislocation slip, with $1/2\langle 111\rangle$ dislocations being the dominant type. When the strain rate increases to $10^{10}$ s$^{-1}$, the yield strength of the alloy is significantly enhanced, accompanied by a notable increase in the proportion of amorphous or disordered structures, indicating that high strain rate loading suppresses dislocation nucleation and motion while promoting disordering transitions. Simulations at varying temperatures indicate that the alloy maintains a high strength level even at temperatures as high as 2100 K. Compositional analysis further indicates that increasing the atomic percentage of Nb or Ta effectively enhances the yield strength of the alloy, whereas an increase in Ti or Zr content adversely affects the strength. By combining MLP with MD methods, this study elucidates the anisotropic characteristics of the mechanical behavior and the strain rate dependence of disordering transitions in the NbTaTiZr RMPEA under combination of high strain rate and high temperature, providing an important theoretical basis and simulation foundation for optimizing and designing novel material under extreme environments.

PACS：61.66.Dk, 02.70.Ns

**doi:** 10.7498/aps.74.20250738

**cstr:** 32037.14.aps.74.20250738

# 1. Introduction

As an important breakthrough in the field of materials science in recent years, multi-principal component alloys have greatly expanded the composition design space by introducing four or more major elements (equimolar/non-equimolar ratio), providing a new paradigm for the development of high-performance alloy systems. Different from traditional alloys, the random mixing of elements in multi-principal component alloys brings high configurational entropy, which can inhibit the formation of intermetallic compounds and facilitate the formation of simple solid solutions with some simple crystal structures, such as face centered cubic (FCC) and body centered cubic (BCC). Among them, the refractory multi-main element alloy containing high melting point elements such as Nb, Mo, Ta, Ti, W, V, Hf, Zr and[1–4] mostly has BCC phase structure, has high melting point, yield strength, creep resistance, thermal stability and energy release characteristics, can still maintain excellent mechanical properties and[5–8] at high temperature, and is expected to be widely used in high temperature environment to promote the rapid development of aerospace and military equipment and other fields[9].

NbTaTiZr refractory multi-principal component alloy has become a hot research topic because of its composition characteristics and performance advantages. The system significantly improves the material properties through process optimization[10,11], and the strength and plasticity prediction based on first-principles calculation provides theoretical support for component design[12]. The experimental results show that the plastic deformability of the alloy is greatly improved by precipitate-induced transformation under quasi-static compression, and the Hopkinson bar experiment further reveals that the alloy has a unique dynamic mechanical response with high strength, high plasticity and low adiabatic shear sensitivity at high strain rates[14]. More breakthrough is that Nb, Ta, Ti, Zr elements show special application potential in the field of energetic materials due to their synergistic oxidation and energy release effect under high velocity impact. Wang et al.[15] and Lo et al.[16] found that the rapid diffusion of oxygen atoms at grain boundaries will form non-protective composite oxides, which provides a new way for energy release. The research on NbTaTiZr system is more focused on experimental research, and the traditional experimental methods are difficult to analyze the nanoscale structural evolution during dynamic loading in real time, and the lack of molecular dynamics simulation of this system restricts the deep optimization of material properties and engineering application development.

Molecular dynamics simulation provides an important means to reveal the atomic-scale mechanical behavior of materials.[17,18], Li et al.[4]studied the strengthening mechanism of NbMoTaW by machine learning potential energy surface model, and found that the precipitation of Nb atoms at grain boundaries is conducive to the strengthening of the alloy; Zhou et al. Constructed a machine learning potential energy surface model of TiZrHfNbO system[19], and found that the oxygen complex formed after adding O atoms significantly increased the critical shear stress required for continuous dislocation motion, which revealed the mechanism of the oxygen complex improving the strength and plasticity of the alloy at the same time, and made up for the shortcomings of experimental observation. However, most of the existing studies are limited to empirical potential functions, which are difficult to accurately describe the complex interactions of multi-principal component systems[20–22]. In recent years, the potential energy surface model based on machine learning has provided a new way to break through the contradiction between accuracy and efficiency. By fitting the first-principle data, the accuracy of the results calculated by density functional theory (DFT) can be achieved while ensuring the computational efficiency[4,19,23–29].

In this paper, the effects of crystal orientation, temperature, strain rate and composition on the compressive mechanical behavior of NbTaTiZr quaternary alloy were systematically studied by developing the potential energy surface model of NbTaTiZr quaternary alloy. The research results will increase our understanding of the characteristics of NbtaTiZr alloy and accelerate the development of high-performance alloys.

## 2. Research method

### 2.1 Deep potential model

In this study, DP-GEN (deep potential GENerator)[30] software was used to explore the configuration space of multi-component alloys and construct the deep potential (DP) model, which has been widely used in different systems[31–34]. The initial structure includes Ta/Ti/Zr/Nb and its binary, ternary and quaternary alloy structures with equal molar ratio. Each component includes three phase structures of BCC/FCC/HCC. For the quaternary alloy, the disordered structure is also included (the disordered structure is obtained by raising to 4000 K and then rapidly cooling to 300 K). The alloy structure is obtained by randomly replacing atoms from a simple substance, and the atoms are randomly distributed in the lattice. The schematic diagrams of the quaternary alloy BCC, FCC, and HCP structures are shown in the Fig. 1 (each supercell structure contains 32 atoms). The initial data set was sampled using three deformation strategies: 1) 0.1 Å atomic displacement perturbation was randomly applied to the 32-atom supercell and ± 20% random perturbation was applied to the box; 2) The structures of different volumes and thermal perturbations were collected in the range of 300 — 900 K and – 5 — 20 GPa; 3) Different thermally disturbed structures (300-900 K) were collected in the NVT ensemble by simultaneously changing the box lengths in three directions (± 1%, ± 2%, ± 3%, ± 4%, ± 5%).

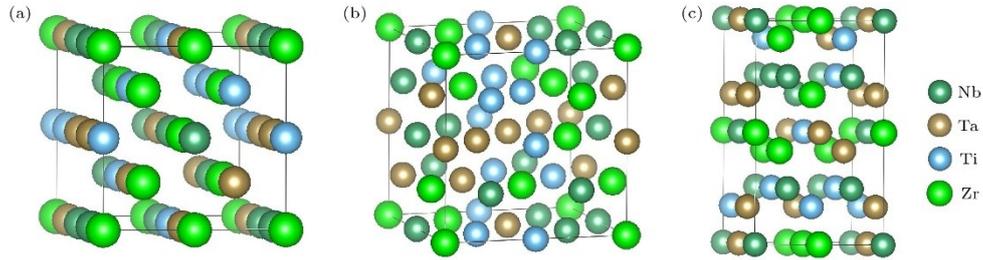

**Figure 1.** Schematic of 32-atom supercell: (a) BCC; (b) FCC; (c) HCP structure.

The DP-GEN iteration is implemented in three stages: first to converge elemental, equimolar binary, ternary alloys, then to extend to non-equimolar binary, ternary systems, and finally to explore quaternary alloys. Each component contains three phase structures, with a total of 14085 initial configurations (4 simple substances, 558 binary, 5580 ternary, and 13485 quaternary). The temperature range explored is 300 — 4000 K, the temperature sampling interval is 300 K in the temperature range of 300 — 2100 K, and the temperature sampling interval is 400 K in the temperature range of 2400 — 4000 K. The pressure sampling range is – 5 — 40 GPa. In the – 5 — 5 GPa range, the pressure sampling interval is 1 GPa, while in the 5 — 40 GPa range, the pressure sampling interval is 5 GPa. In order to ensure that more representative configurations can be collected in each iteration, the Monte Carlo and molecular dynamics method is used to sample, which can collect the structures of different elements with the same element ratio; At the same time, one structure is output at an interval of 0.5 ps for each dynamic trajectory (20 ps) to ensure that there is a large difference between the two output structures; In addition, the confidence interval of force deviation is dynamically adjusted in the iterative process to ensure that the marked structure has a large difference (in the DP-GEN iterative process, the force of the structure output by the dynamic trajectory will be calculated. If the maximum difference between the force values of the four DP models for the same configuration is greater than the upper limit of the confidence interval, the configuration is judged to be a failure; The configuration below the lower limit of the confidence interval is judged to be an accurate configuration, that is, the current model has the ability to describe the configuration without being added to the training set; Only configurations that fall within the confidence interval may be labeled and added to the training set). The DeepMD-kit[35] software package is used to fit the DFT data, and the se _ attent[36] descriptor is selected, which has high prediction accuracy in the multi-element system and can reduce the computational and training costs. The truncation radius is 6 Å, the embedding network is (25,50,100), the fitting network is (240,240,240), the weight factors of the loss function are energy (0.02,1), force (1000,1) and virial stress (0.02,1), respectively, and the learning rate decays from the initial 0.001 to $1 \times 10^{-8}$ through $2 \times 10^5$ steps. When the accurate configuration reaches 95%, the iteration is considered to be convergent.

The VASP (Vienna *ab* initio simulation package) [37–38] initio simulation package package was used to calculate the energy, force and virial stress of the initial data set and candidate structures. The kinetic energy of the plane wave basis set was truncated at 500 eV, and the convergence criterion of the electronic structure optimization was $10^{-7}$ eV. According to the Monkhorst-Pack method, the grid density of $k$ points in the Brillouin zone is 0.2 Å$^{-1}$, and the PBE functional under the generalized gradient approximation (GGA)[39,40] is used to describe the exchange-correlation interaction. The elastic constants of the alloy were calculated by vaspkit[41] software package. In the calculation of the structural elastic constants, the $k$ point density, the kinetic energy cutoff of the plane wave basis set and the electronic structure convergence criterion were consistent with the DFT calculation parameters when the data set was constructed.

## 2.2 MD simulation setup

Firstly, the BCC supercell structure of Ta was established, and then Ta atoms were randomly replaced by Nb, Ti and Zr atoms according to their proportions. It is found that the compression results of the initial structure obtained by replacing atoms with different random numbers are consistent, so the influence of random atom occupancy is not considered when discussing the influence of different factors on the compression mechanical properties of the alloy. The expansion coefficients along [100], [110] and [111] directions are 20×20×40, 12×12×24, and 12×12×24, respectively, and the total number of atoms are 32000, 27648 and 27648. Before the structure is compressed, it is first energy minimized and then relaxed at NPT (300 K, 0 GPa) ensemble for 50 PS. The compression process was also carried out in NPT ensemble at 300 K, with a $x$ perpendicular to the compression direction and a pressure of 0 in the $y$ direction. The atomic structure was analyzed by OVITO[42] software, and the crystal defects were characterized by common neighbor analysis[43] and dislocation analysis[44] algorithms.

## 3. Results and Discussion

### 3.1 Potential function verification

The Fig. 2(a) —(c) shows the distribution of energy, force and virial stress of the training set. Because different temperature and pressure intervals, different alloy components and different structures are sampled, the structure itself has differences in physical quantities such as energy, force and virial stress, and the data set is unevenly distributed in different intervals due to the randomness of the marked structure in the iterative process. The Fig. 2(d) and(f) show the prediction of energy, force and virial stress of the trained DP model on the whole training set and test set. The root mean square errors of energy, force and virial stress on the training set are 13.2 meV/atom, 202.7 meV/Å and 40.6 meV/atom, respectively, indicating that the obtained DP model has high fitting accuracy; On the test set, the root mean square errors of energy, force and virial stress are 13.99 meV/atom, 228.71 meV/Å and 44.18 meV/atom, respectively, which indicates that the DP model has high prediction accuracy.

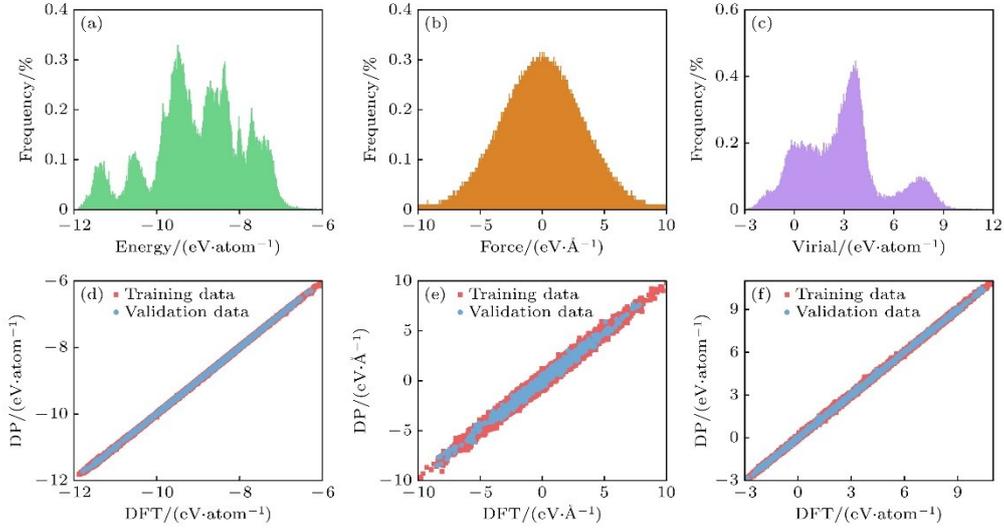

**Figure 2.** (a) Atomic energy, (b) forces, (c) distribution of virial stress; correlation between DP and DFT for (d) energy, (e) forces and (f) virial stress.

The accuracy of the potential function model is further verified by calculating the elastic constants and melting point of the alloy, and the results are shown in Tab. 1. DP and DFT calculations are based on the same structure. The results show that the DP calculations are in good agreement with the experimental measurements (melting point) and the DFT calculations. Except for the shear modulus, the relative errors of the other calculations (the ratio of the absolute value of the difference between DP and DFT calculations to DFT) are less than 10%.

Table 1. Comparison of DP model predictions with DFT calculations and experimental values.

|  | Method | $T_m$/K | $C_{11}$/GPa | $C_{12}$/GPa | $C_{44}$/GPa | $B$/GPa | $G$/GPa | $\mu$ |
|---|---|---|---|---|---|---|---|---|
| NbTaTiZr | DP | 2361 | 170 | 129 | 30 | 142 | 32 | 0.43 |
|  | Exp/DFT | 2440 (3.2%) | 179 (5.0%) | 134 (3.7%) | 33 (10%) | 151 (6.0%) | 27 (18.5%) | 0.42 (2.4%) |

## 3.2 Compressive mechanical property

The compressive stress-strain curve of NbTaTiZr alloy along [100] grain direction is shown in Fig. 3(a). In the initial stage, the compressive stress increases almost linearly with the increase of strain. When the strain increases to about 15.78%, the stress decreases sharply. This phenomenon of stress decreasing sharply with the increase of strain after reaching the peak value is common in molecular dynamics simulation[45–47]. The mechanism is that the elastic strain energy accumulated in the elastic deformation stage is released at this time, resulting in the decrease of stress. The peak compressive stress of the stress-strain curve is defined as the yield strength of the material, which is 10.8 GPa. After the stress drops to the lowest point, as the strain continues to increase, the curve shows a slight rebound with a slight fluctuation.

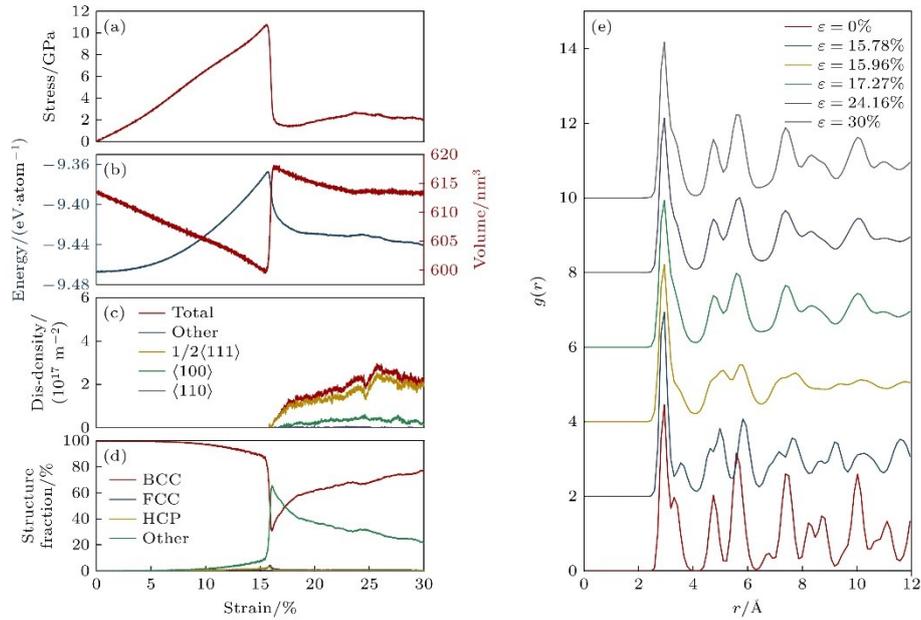

**Figure 3.** Evolution of properties for NbTaTiZr under uniaxial compression along the [100] crystalline orientation as a function of compressive strain at 300 K and a strain rate of $10^9$ s$^{-1}$: (a) Compressive stress; (b) energy and volume; (c) dislocation density; (d) phase structure. (e) Radial distribution functions at different strain levels.

Fig. 3(b) is the variation of volume and monatomic potential energy during compression. In the elastic stage, the energy of the system increases due to the work done by the outside world, and the point of sudden decrease of stress corresponds to the point of sudden decrease of energy, which indicates that the release of elastic strain energy reduces the total energy of the system; The volume change curve shows that the volume of the system decreases with the increase of strain at the initial stage of compression, while the volume increases suddenly near the point of sudden stress drop, indicating that the structure expands and stress relaxes. The evolution of dislocation density in the whole compression process is shown in the Fig. 3(c). The main dislocation type in the system is 1/2 ⟨111⟩, followed by a small amount of ⟨100⟩ dislocation, which is consistent with the research trend of Zhang's[48]. Dislocation begins to nucleate when the strain reaches 15.78%, and then the dislocation density increases gradually. The change trend of dislocation density is consistent with the change law of the stress-strain curve after a sudden drop: the dislocation density increases with the increase of stress and decreases with the decrease of stress, which indicates that dislocation slip is an important mechanism of plastic deformation in the alloy with grain direction down. The Fig. 3(d) shows the change of phase structure during compression, and a small amount of BCC structure has transformed into disordered structure in the elastic stage. In the strain range of 15.8% — 16.19%, the proportion of BCC phase structure decreases sharply, while the proportion of disordered structure increases sharply, which is highly consistent with the stress drop range in the Fig. 3(a), indicating that the transformation from BCC phase to disordered structure is the main reason for the significant stress drop. After that, the proportion of BCC phase structure gradually rises and the proportion of disordered structure decreases, which is consistent with the change trend of the stress-strain curve. In order to explore the microstructure evolution of the alloy, Fig. 3(e) shows the radial distribution function (RDF) of the structure under six different strain conditions. When the strain is 15.96%, the peak value of the RDF curve is smaller than that under other strains, and the overall RDF curve is more gentle, indicating that a large number of atoms in the system are in a disordered state, while the RDF curves under other strains have multiple sharp peaks, indicating that there are a large number of crystal ordered structures in the corresponding structure.

Further analysis of the microstructure evolution, Fig. 4 shows the dislocation morphology and phase structure distribution of NbTaTiZr alloy under different strains. Dislocation lines are colored according to type, where green replaces Tab. 1/2 $\langle 111 \rangle$ dislocation, red represents $\langle 100 \rangle$ dislocation, and the dislocation begins to nucleate at strain of 15.78%. With the increase of strain, the dislocation density increases gradually, and the length of dislocation line also increases correspondingly, which is consistent with the variation trend of dislocation density in Fig. 3(c). The spatial distribution of phase structure is clearly shown in Fig. 4(a2) —(e2). At low strain, disordered structure can be observed in some regions. When the strain increases to 15.96%, a large and relatively uniform disordered structure region is formed in the system. Then, with the increase of strain, the proportion of disordered atoms gradually decreases and gradually transforms into BCC structure.

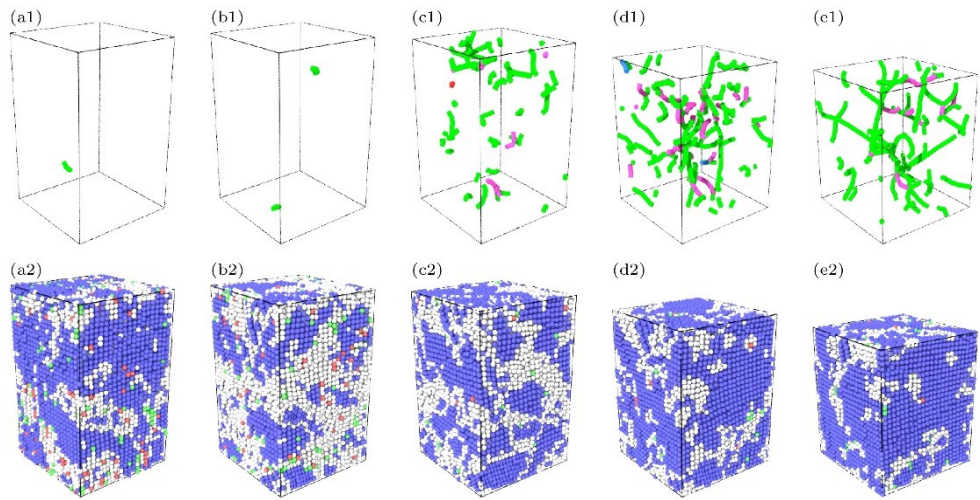

**Figure 4.** Dislocation density (a1)–(e1) and phase structure distribution (a2)–(e2) in NbTaTiZr under uniaxial compression along the [100] orientation at 300 K and a strain rate of $10^9$ s$^{-1}$, shown at compressive strains of 15.78%, 15.96%, 17.27%, 24.16%, and 30%. Blue represents BCC, red represents HCP, green represents FCC, and grey represents other structures.

## 3.3 Effect of crystal orientation on compressive mechanical properties

The compressive mechanical behavior of NbTaTiZr alloys usually exhibits a significant orientation dependent[45]. A thorough understanding of this orientation dependent mechanical response is of great significance for a comprehensive understanding of the compression deformation mechanism of NbTaTiZr alloys and the design of advanced alloys. For this reason, we calculated the compressive behavior of the alloy along [110] and [111] crystallographic directions, and the stress-strain curves are shown as Fig. 5(a),(b). Combined with the comparative analysis of the results of [100] crystal direction in the Fig. 3(a), it can be clearly observed that the stress-strain curves show obvious anisotropy. Specifically, when compressed along [111] and [100] crystal directions, the stress-strain curves show similar variation laws: the stress first increases nearly linearly to the peak value, and then decreases suddenly, while when compressed along [110] crystal direction, the fluctuation of the stress state is significantly smaller than that of the other two directions. The yield strength of compression along [110] crystal direction is the lowest, while that along [. The Fig. 5(c) and(d) show the evolution of the compressed structure with strain along different crystallographic directions. Combined with the Fig. 3(d) analysis, the structural change along [100] and [111] crystallographic directions is similar, and the disordered structure has a very obvious mutation point, and the maximum proportion of the disordered structure is higher than that of the BCC structure under the same strain, which also corresponds to the mutation point of the stress-strain curve. However, the content of disordered structure in [110] direction is always lower than that of BCC phase, and there is no obvious increase or decrease point. After the structure reaches yield, the proportion of disordered structure fluctuates in a small range, which is consistent with the change trend of the stress-strain curve.

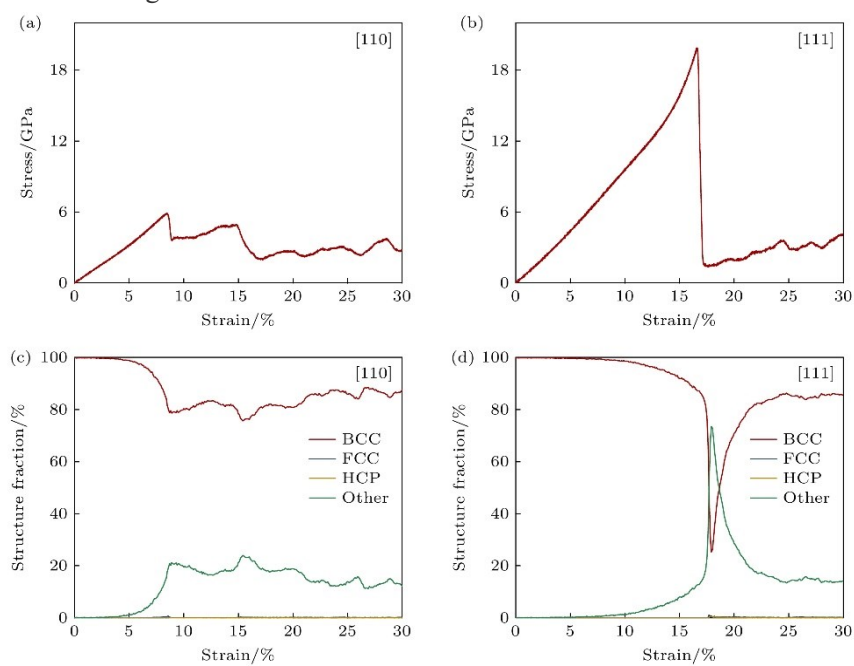

**Figure 5.** Stress-strain curves (a), (b) and structural evolution (c), (d) under uniaxial compression along the different orientations: (a), (c) [110] orientations; (b), (d) [111] orientations.

From the Fig. 5(c), it can be found that the content of disordered structure is not only the lowest when compressed along [110] direction, but also its variation trend is significantly different from that along [100] and [111] directions, which indicates that the deformation mechanism of [110] direction may be different from that of [111] and [100] directions. In order to explore the microscopic mechanism, the Fig. 6(a1) and(a4) show the microstructure compressed to different strains along the [110] crystal direction at 300 K. Before the yield point, there are only sporadic disordered regions in the structure. With the increase of strain, these disordered structural sites undergo large shear deformation, which makes the deformation more localized. At the same time, deformation twins begin to nucleate and grow, and the twin boundary undergoes propagation, proliferation, and annihilation, as shown in the inset of the Fig. 6(a2) —(a4). The fluctuation of the proportion of disordered atoms in the structure leads to the fluctuation of the stress-strain curve after the yield point. The Fig. 6(b1) —(b4) shows the local shear strain distribution under different strains, and the analysis shows that the larger shear strain is highly consistent with the twin contour, which confirms the dominant role of twin deformation in the plastic deformation.

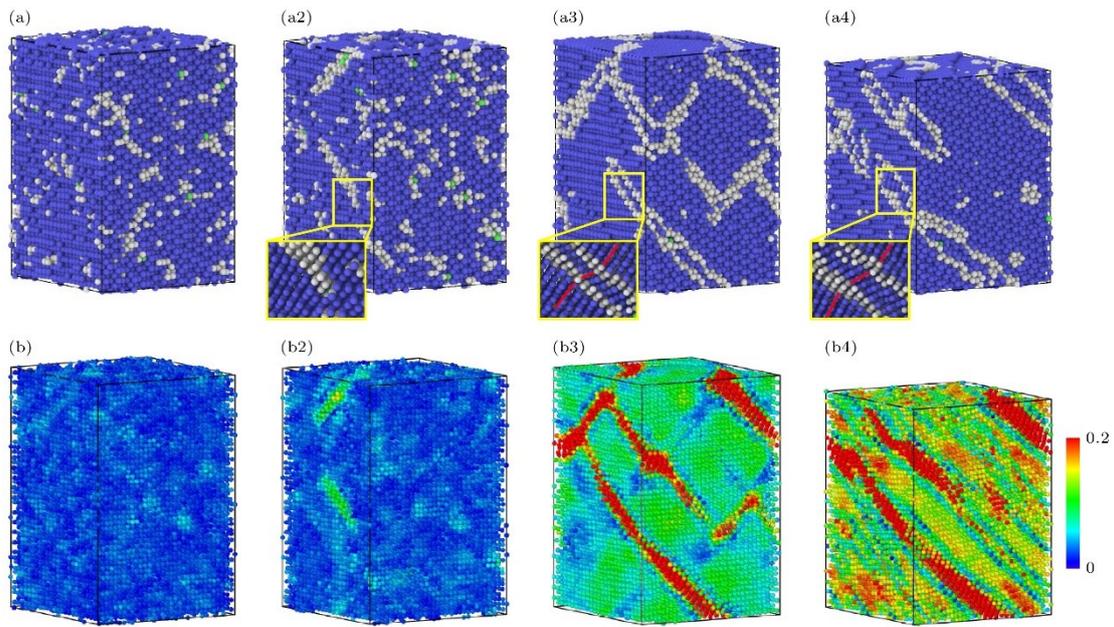

**Figure 6.** Atomic structures (a1)—(a4) and shear strain distribution (b1)—(b4) in NbTaTiZr under uniaxial compression along the [110] orientation at 300 K and $10^9$ s$^{-1}$ for varying strains: (a1), (b1) 8.3%; (a2), (b2) 8.6%; (a3), (b3) 8.9%; (a4), (b4) 14.9%.

Similar to the results of compression along [100] direction, the deformation mechanism of compression along [111] direction is also dominated by dislocation slip and local disordering, and the distribution of dislocation configuration and phase structure under different strains is shown in Fig. 7. When the strain is in the range of 16.7% — 17.4% (Fig. 7(a1),(b1)), the dislocation lines are short and scattered. When the strain increases to 18.7% (Fig. 7(c1)), the dislocation lines become long. After that, with the increase of strain, the dislocations are annihilated and the density of dislocation lines decreases. When the stress state suddenly decreases from the yield point, there is an obvious disorder transformation (Fig. 7(b2)). After yielding, most of the disordered structure transforms into BCC structure, and a small number of disordered atoms are scattered.

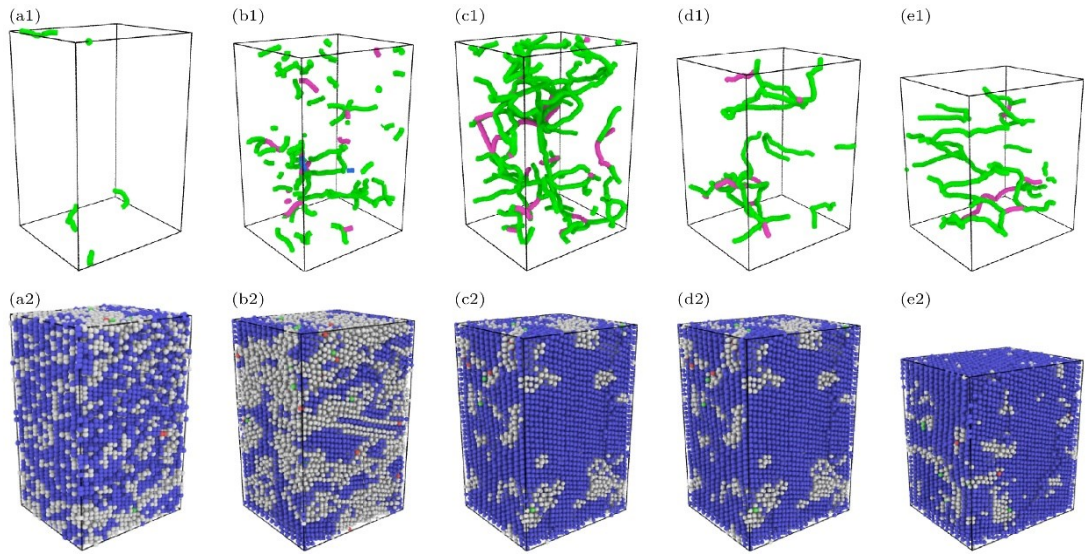

**Figure 7.** Dislocation density (a1)–(e1) and phase structure distribution (a2)–(e2) in NbTaTiZr under uniaxial compression along the [111] orientation at 300 K and a strain rate of $10^9$ s$^{-1}$ for varying strains, and blue represents BCC, red represents HCP, green represents FCC, and grey represents other structures: (a1), (a2) 16.7%; (b1), (b2) 17.4%; (c1), (c2) 18.7%; (d1), (d2) 24%; (e1), (e2) 30%.

Combined with the results of 3.2 section, it can be found that the compressive mechanical properties of NbTaTiZr alloy along different crystal directions have significant anisotropy. The compressive yield strength along [110] direction is the lowest, the compressive strength along [111] direction is the highest, and the compressive yield strength along [100] direction is in the middle. This may be related to the interatomic spacing in different crystal directions, with the interatomic spacing in [110] direction being the largest, that in [111] direction being the smallest, and that in [100] direction being intermediate. The larger the atomic spacing, the weaker the interatomic interaction, while the smaller the atomic spacing, the stronger the interatomic interaction, thus showing a higher yield strength, which also shows a similar rule in BCC metal Mo[49].

## 3.4 Effect of strain rate on compressive mechanical properties

Strain rate is a key factor affecting the compressive mechanical behavior of materials,[50–52]. To study its effect, the compressive mechanical behavior of NbTaTiZr alloy was calculated at strain rates lower than ($10^8$ s$^{-1}$) and higher than ($10^{10}$ s$^{-1}$) $10^9$ s$^{-1}$, and the corresponding stress-strain curves are shown in Fig. 8(a),(b). With the increase of strain rate, the yield strength of the alloy increases gradually, and the yield strength is 10. 2 GPa at $10^8$ s$^{-1}$ and 13. 8 GPa at $10^{10}$ s$^{-1}$. The linear sections of the stress-strain curves at the three strain rates almost coincide, that is, the elastic modulus is almost not affected by the strain rate. Increasing the strain rate can improve the yield strength of the alloy, which is also found in $Co_{25}Ni_{25}Fe_{25}Al_{7.5}Cu_{17.5}$[50], FeNiCrCoCu[51], FeNiCrCoAl[53] and other systems. When the stress reaches the yield point, the stress-strain curve shows a sudden downward trend as the strain continues to increase. During the subsequent loading process, serrated flow occurs at low strain rates ($10^8$ s$^{-1}$), but not at high strain rates ($10^{10}$ s$^{-1}$); The previous study shows that 1/2 ⟨111⟩ is the most important dislocation type in the alloy during compression. Fig. 8(b) shows the change of 1/2 ⟨111⟩ dislocation density under different strain rates. The dislocation density under $10^8$ s$^{-1}$ and $10^9$ s$^{-1}$ loading conditions is close, and when the strain rate increases to $10^{10}$ s$^{-1}$, the dislocation almost disappears.

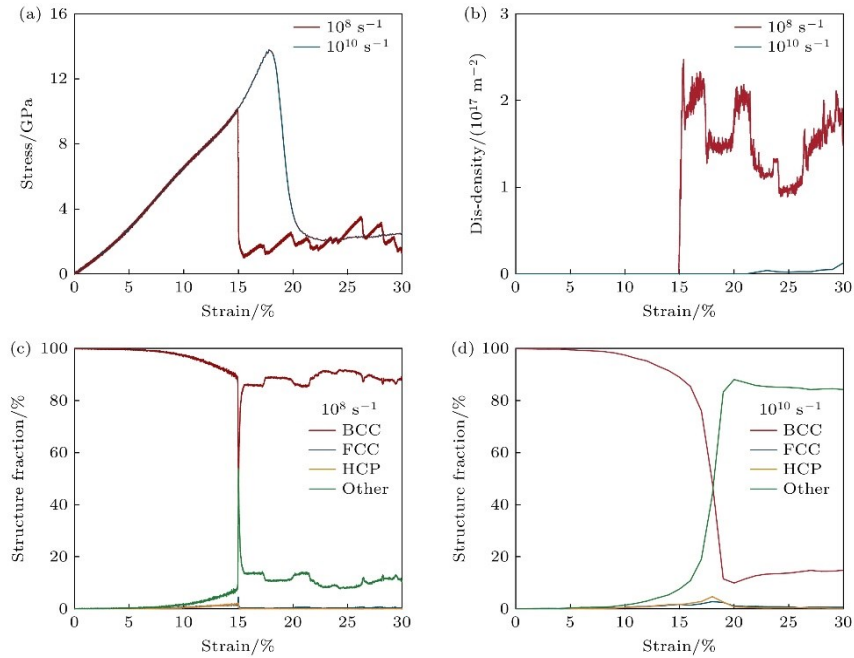

**Figure 8.** (a), (b) Stress-strain curves and dislocation density at strain rates of $10^8$ s$^{-1}$ and $10^{10}$ s$^{-1}$; (c), (d) phase proportion versus strain at $10^8$ s$^{-1}$ and $10^{10}$ s$^{-1}$.

Further analysis of the phase structure change during compression shows that the disordered structure in the alloy will suddenly increase after the structure is yielded, and only under the strain rate of $10^{10}$ s$^{-1}$, the disordered structure will remain in a high proportion. Under the other two strain rates, the disordered structure will transform into BCC structure again with the increase of strain. The Fig. 9(a) shows the yield strength of [100], [110] and [111] crystallographic directions at different strain rates. Under all strain rates, the yield strength along [111] direction is always the highest, while the yield strength along [110] direction is always the lowest. In order to quantitatively characterize the degree of disordering, the average value of the proportion of disordered atoms in the strain range of 20% — 30% was taken as the proportion of disordered structure at the strain rate, and the Fig. 9(b) was the content of disordered structure along three different crystal directions at three different strain rates. It can be seen that the proportion of disordered structure increases with the increase of strain rate, which indicates that high strain rate significantly promotes the transformation of the alloy to disordered state[50]. In the range of $10^9$ to $10^{10}$ s$^{-1}$, the yield strength and the content of disordered structure increase significantly faster than in the low strain rate range ($10^8$ to $10^9$ s$^{-1}$). The evolution of 1/2⟨111⟩ dislocation density under loading strain rates of $10^8$ s$^{-1}$ and $10^{10}$ s$^{-1}$ is shown in Figures 9(c)–(d). The dislocation density also exhibits obvious anisotropy and strain rate sensitivity. Under the high condition of $10^{10}$ s$^{-1}$, dislocations still exist in the [110] orientation, though they are reduced compared to the structure at a strain rate of $10^8$ s$^{-1}$. In contrast, the compression results for the [111] orientation are similar to those for the [100] orientation, with almost no 1/2⟨111⟩-type dislocations present at a strain rate of $10^{10}$ s$^{-1}$.

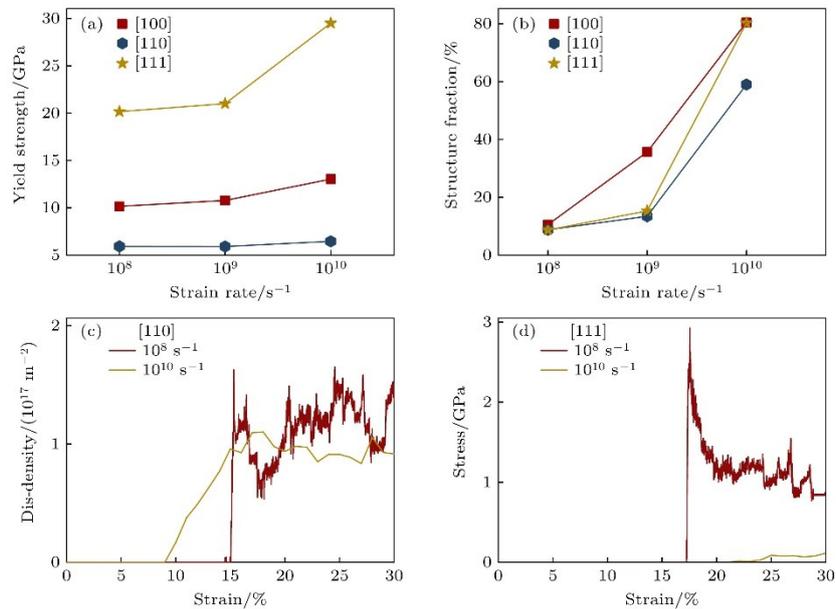

**Figure 9.** Yield strength (a) and disordered structure proportion (b) under [100], [110], and [111] crystal orientations at strain rates of $10^8$ s$^{-1}$, $10^9$ s$^{-1}$, and $10^{10}$ s$^{-1}$; dislocation density versus strain for [110] orientation (c) and [111] orientation (d) at strain rates of $10^8$ s$^{-1}$ and $10^{10}$ s$^{-1}$.

The Fig. 10 shows the results of different crystal directions and strain rates through the atomic arrangement and dislocation line distribution: at lower strain rates ($10^8$ s$^{-1}$), the proportion of disordered atoms in the three crystal directions is low, and the disordered atoms are mainly attached to the vicinity of dislocation lines, while the regions far away from dislocation lines are relatively small. However, when the strain rate is increased to $10^{10}$ s$^{-1}$, the disordered structure under the three crystal directions increases greatly, and the disordered atoms not only attach to the dislocation lines, but also widely exist in the dislocation-free region. This indicates that the deformation mechanism changes at high strain rates: high strain rates inhibit the formation and proliferation of dislocations, thus promoting the formation of non-dislocation-dominated, more uniform disordered structures.

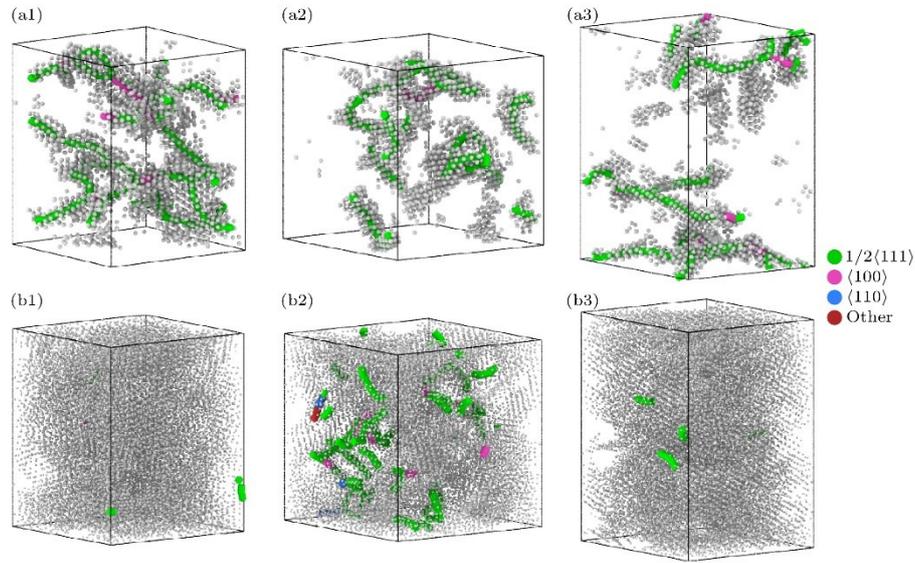

**Figure 10.** Dislocation configurations at 28% strain under strain rates of $10^8$ s$^{-1}$ (a1)–(a3) and $10^{10}$ s$^{-1}$ (b1)–(b3), and the gray atoms indicate disordered atomic configurations: (a1), (b1) [100] crystallographic orientations; (a2), (b2) [110] crystallographic orientations; (a3), (b3) [111] crystallographic orientations.

## 3.5 Effect of Composition on Compressive Mechanical Properties

Adjusting the composition has a significant effect on the mechanical properties of the alloy[54]. The compressive mechanical responses of the alloy were calculated at 300 K and $10^9$ s$^{-1}$ with the increase and decrease of Nb, Ta, Ti and Zr contents, respectively. The increase or decrease of the element content is 10% relative to the equimolar system (each element content is 25%). When one element is added, the total content of the other three elements decreases, but the molar ratio remains equal. In order to simplify the form, the chemical formulas of each individually reduced or increased element content are written as $Nb_{0.15}$, $Nb_{0.35}$, $Ta_{0.15}$, $Ta_{0.35}$, $Ti_{0.15}$, $Ti_{0.35}$, $Zr_{0.15}$, $Zr_{0.35}$ in turn, representing the chemical formulas of Nb, Ta, Ti, Zr elements reduced or increased by 10% in turn. The corresponding stress-strain curve is shown in Fig. 11. In all cases, the stress increases linearly with the increase of strain. After reaching the yield point, the elastic strain energy is suddenly released, resulting in a sudden decrease in stress, followed by plastic deformation. The increase of Nb or Ta content can significantly improve the yield strength and elastic modulus of the alloy, while the increase of Ti or Zr content has the opposite effect.

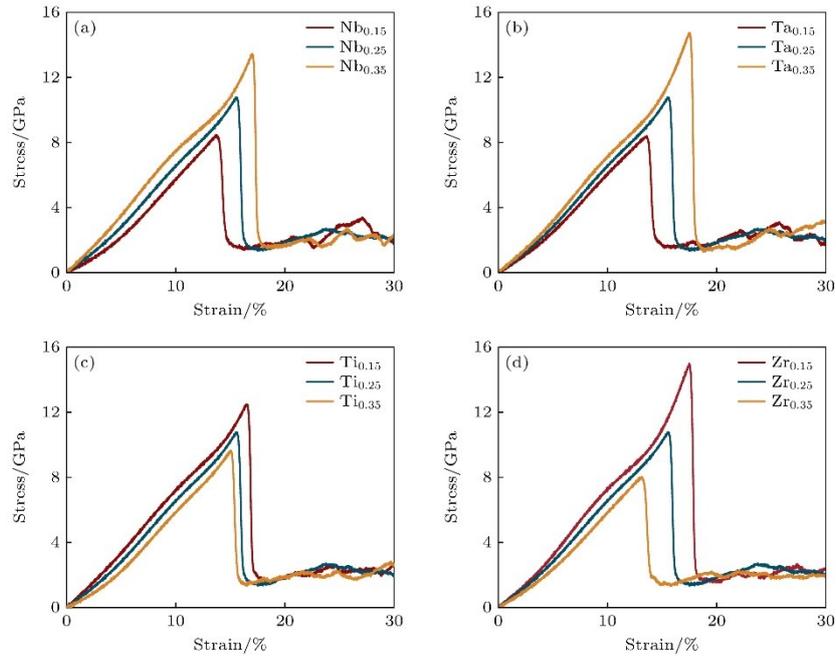

**Figure 11.** Compressive stress-strain curves for different content of element: (a) Nb, (b) Ta, (c) Ti, (d) Zr.

In order to systematically study the compositional effect and the variation of yield strength of the alloy at different temperatures, the compressive mechanical response along the [100] crystal direction in the range of 300 — 2100 K (300 K, 900 K, 1500 K, 2100 K) was additionally calculated. Similarly, when the proportion of a certain element is changed, the other three elements are kept in an equal molar ratio, and the content of the element is 10%, 30%, 50%, 70% and 90% in turn, and the total content of the other three elements is 90%, 70%, 50%, 30% and 10%. Fig. 12 is the yield strength nephogram of different components. It can be found that with the increase of temperature, the yield strength of alloys with all components will decrease. This is because the increase of temperature leads to the intensification of atomic thermal motion and the increase of atomic migration probability, which reduces the yield strength of alloys. In the whole composition space variation, the yield strength of the alloy can be improved by increasing the Nb content and keeping the equal molar ratio of Ta, Ti and Zr, and the same is true by increasing the Ta content alone. At high temperature, the effect of increasing Ta content on yield strength is slightly stronger than that of increasing Nb content. However, increasing the Ti content and keeping the equimolar ratio of Nb, Ta and Zr will reduce the yield strength of the alloy, which is the same as increasing the Zr content alone. The effect of simultaneously changing the ratio of multiple elements on the compressive mechanical properties of the alloy will be studied in the future.

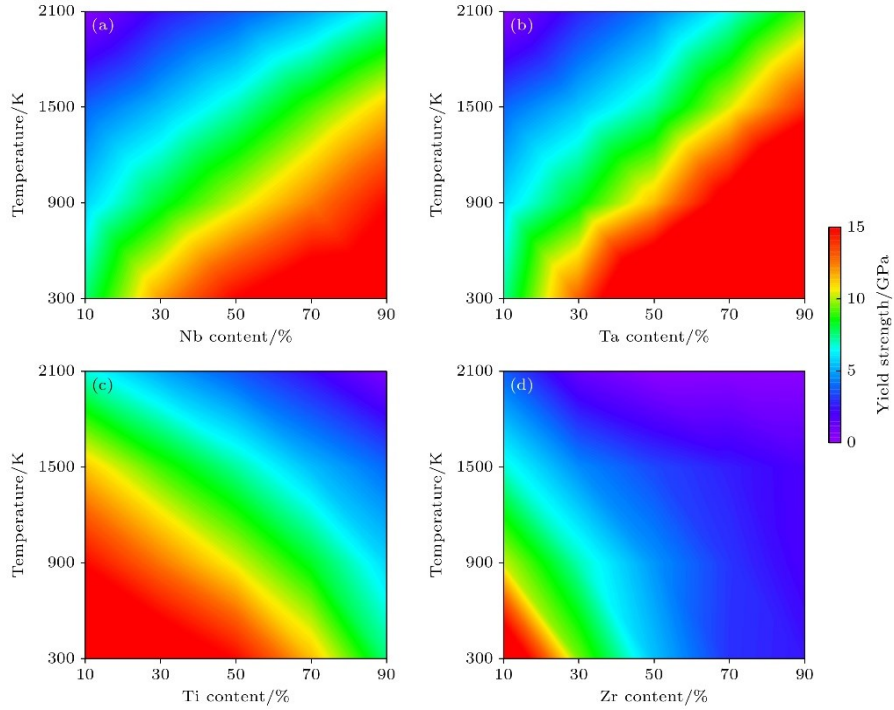

**Figure 12.** Effects of varying the content of a single element on yield strength across the 300—2100 K temperature range: (a) Nb; (b) Ta; (c) Ti; (d) Zr.

## 4. Conclusion

In this study, the deformation mechanism and mechanical response of NbTaTiZr quaternary refractory multi-principal component alloy under compression were studied by constructing a high-precision machine learning potential function. It was found that under the conditions of $10^9$ s$^{-1}$ loading strain rate and 300 K, the system released elastic strain energy through the transformation from BCC phase to disordered state when compressed along the [100] crystal direction, and the dislocation was mainly 1/2 ⟨111⟩ type; Further analysis shows that the mechanical behavior is significantly anisotropic, the [111] crystal direction shows the highest yield strength, while the [110] crystal direction shows the lowest, and the structure is coordinated by deformation twins when compressed along the [110] crystal direction. It is also found that when the strain rate increases from $10^8$ s$^{-1}$ to $10^{10}$ s$^{-1}$, the yield strength increases significantly, and the proportion of disordered structure in the structure also increases significantly, which confirms that high strain rate promotes the disorder transformation by inhibiting dislocation movement. It is worth noting that the alloy still maintains excellent strength at 2100 K, and the increase of Nb/Ta content can significantly improve the yield strength of the alloy, but the increase of Ti/Zr content has a negative effect.